\documentclass[onecolumn]{aastex631}
\usepackage{amsmath}
\usepackage{diagbox}
\usepackage{graphicx}
\usepackage{makecell}
\usepackage{enumitem}

\received{}
\revised{}
\accepted{}

\submitjournal{AAS}

\shorttitle{Estimating luminosity functions via KDE}
\shortauthors{Yuan et al.}

\begin{document}

\title{A flexible method for estimating luminosity functions via Kernel Density Estimation - III. Extending to Multiple Flux-Limited Samples}

\correspondingauthor{Zunli Yuan}
\email{yzl@hunnu.edu.cn}

\author[0000-0001-6861-0022]{Zunli Yuan}
\affiliation{Department of Physics, School of Physics and Electronics, Hunan Normal University, Changsha 410081, People's Republic of China}
\affiliation{Key Laboratory of Low-dimensional Quantum Structures and Quantum Control, Hunan Normal University, Changsha 410081, People's Republic of China}
\affiliation{Hunan Research Center of the Basic Discipline for Quantum Effects and Quantum Technologies, Hunan Normal University, Changsha 410081, People's Republic of China}

\author{Chuanqi Li}
\affiliation{Department of Physics, School of Physics and Electronics, Hunan Normal University, Changsha 410081, People's Republic of China}
\affiliation{Key Laboratory of Low-dimensional Quantum Structures and Quantum Control, Hunan Normal University, Changsha 410081, People's Republic of China}
\affiliation{Hunan Research Center of the Basic Discipline for Quantum Effects and Quantum Technologies, Hunan Normal University, Changsha 410081, People's Republic of China}

\author[0009-0005-1617-2442]{Wenjie Wang}
\affiliation{Department of Physics, School of Physics and Electronics, Hunan Normal University, Changsha 410081, People's Republic of China}
\affiliation{Key Laboratory of Low-dimensional Quantum Structures and Quantum Control, Hunan Normal University, Changsha 410081, People's Republic of China}
\affiliation{Hunan Research Center of the Basic Discipline for Quantum Effects and Quantum Technologies, Hunan Normal University, Changsha 410081, People's Republic of China}

\author{Luozhenhan Liu}
\affiliation{Department of Physics, School of Physics and Electronics, Hunan Normal University, Changsha 410081, People's Republic of China}
\affiliation{Key Laboratory of Low-dimensional Quantum Structures and Quantum Control, Hunan Normal University, Changsha 410081, People's Republic of China}
\affiliation{Hunan Research Center of the Basic Discipline for Quantum Effects and Quantum Technologies, Hunan Normal University, Changsha 410081, People's Republic of China}

\author{Longhua Qin}
\affiliation{Department of Physics, Yuxi Normal University, Yuxi, Yunnan, 653100, People's Republic of China}




\begin{abstract}
As the third paper in a series regarding the estimation of luminosity functions (LFs) via kernel density estimation (KDE), we present a further generalization of our framework by extending its applicability to multiple flux-limited samples.
While our previous works addressed single flux-limited datasets, many practical applications involve surveys that cover disjoint fields of view with different flux limits. We introduce a piecewise estimation framework that partitions the luminosity-redshift plane into disjoint regions according to the staggered flux limits of the sub-samples. Within each region, we integrate data from all surveys capable of detecting sources into a combined sample and apply the transformation-reflection KDE method using the corresponding local flux threshold as the truncation boundary. This strategy allows for the full utilization of all available sources while maintaining rigorous statistical consistency. The robustness of this approach is validated through Monte Carlo simulations.
Furthermore, application to SDSS DR7 and 2SLAQ quasar data shows overall agreement with parametric models, while a small residual discontinuity near a survey-transition boundary is discussed as a diagnostic of independent piecewise estimation and possible inter-survey systematics.
The KDE calculations in each piecewise region are performed using our previously developed public Python package \texttt{kdeLF}.
\end{abstract}


\keywords{methods: data analysis --- methods: statistical --- galaxies: luminosity function, mass function.}



\section{Introduction}
\label{Intro}
The luminosity function (LF) of galaxies is a fundamental statistical tool used in extragalactic astronomy to study the distribution and evolution of galaxies and active galactic nuclei (AGNs) over cosmic time. Accurately estimating the LF is crucial for understanding the structure and evolution of the universe. However, observational challenges such as selection effects, particularly flux limits, can introduce systematic uncertainties in LF estimates by artificially truncating the observable $(z, L)$ domain. While traditional binned approaches \citep[e.g., the $1/V_{\text{max}}$ estimator of][]{1968ApJ...151..393S} remain a cornerstone of the field, they often rely on fixed discretization, which can be sensitive to the choice of binning and may lead to information loss near the survey limits \citep[e.g.,][]{2013ApSS.345..305Y}.  To address these limitations, nonparametric methods like kernel density estimation (KDE) offer a powerful alternative by replacing discrete bins with smooth, continuous kernels. In the first two papers of this series \citep[hereafter Paper I and Paper II, respectively]{2020ApJS..248....1Y, 2022ApJS..260...10Y}, we established a ``transformation-reflection" KDE framework to mitigate boundary effects by mapping the truncated data into an unbounded space, thereby enabling a robust and flexible reconstruction of the underlying distribution.

While Paper I and Paper II demonstrated that this KDE approach provides accurate and stable LF estimates for samples with a single flux limit, the framework has already gained traction within the astronomical community.
For instance, \citet{2023ApJ...943..162P} employed our adaptive KDE method to derive the intrinsic X-ray LF of AGNs. By bypassing the need for a fixed functional form, they allowed the data to naturally manifest the ``knee" of the distribution (refer to their Figure 15). Crucially, they integrated this non-parametric LF to compute the cosmic evolution of the black hole accretion density. This model-independent estimate provided a robust baseline for comparison with theoretical simulations, demonstrating that Compton-thick AGNs contribute to the accretion history of AGN as much as all other AGN populations combined. Similarly, \citet{2026ApJ...997..176W} utilized our adaptive KDE approach to reconstruct the radio LFs of star-forming galaxies \citep[see also][]{2024A&A...683A.174W}. Their work enabled a continuous reconstruction across both redshift and luminosity without the limitations of binning or parametric assumptions, revealing clear signatures of joint luminosity and density evolution within the star-forming galaxy population.

Despite these successes in handling single-limit datasets, in real-world astronomical surveys, samples often span multiple regions of the sky, each with different flux limits due to varying survey sensitivities or observational depths. These multiple flux limits complicate the process of combining these datasets for a unified LF estimate. For instance, a survey might combine several disjoint sky regions, each with different flux limits. However, our current KDE framework is not yet capable of handling such multi-sample datasets.

To resolve this, it is instructive to draw a parallel with the evolution of traditional estimators. Just as the classic $1/V_{\text{max}}$ method \citep{1968ApJ...151..393S} was historically generalized to the $1/V_a$ estimator \citep{1980ApJ...235..694A} to accommodate multiple flux-limited samples, the KDE approach likewise requires a multi-sample extension to remain a robust and complete non-parametric alternative. In this paper, we fulfill this requirement by extending our KDE approach to handle a collection of multiple flux-limited samples, where each sub-sample is associated with a distinct flux limit function. Our method enables the integration of these heterogeneous datasets while accounting for the different flux thresholds. By generalizing our transformation-reflection KDE framework, we aim to estimate the LF more accurately by fully utilizing the information from all survey regions, thus overcoming the challenges posed by varying flux limits. This advancement is particularly important for surveys covering multiple sky regions or those with varying observational conditions, common in modern extragalactic surveys such as the Sloan Digital Sky Survey (SDSS), the Dark Energy Survey (DES), or the Legacy Survey of Space and Time (LSST), and future wide-field radio surveys like the Square Kilometre Array (SKA).

Throughout the paper, we adopt a Lambda Cold Dark Matter cosmology with the parameters $\Omega_{m}$ = 0.30,  $\Omega_{\Lambda}$ = 0.70, and $H_{0}$ = 70 km s$^{-1}$ Mpc$^{-1}$.

\section{Methodology}
\label{LF_KDE}

\subsection{Kernel Density Estimation}
\label{sec:kde_basics}

Kernel Density Estimation (KDE) is a well-established non-parametric statistical method used to estimate an unknown probability density function (PDF) from a finite data sample. Unlike histograms, which rely on fixed binning, KDE places a smooth kernel function at each data point and sums these kernels to construct a continuous and smooth density estimate.

Since our LF calculation is a two-dimensional (2D) problem (i.e., based on redshift $z$ and luminosity $L$), we use the bivariate KDE estimator. For a set of 2D data points $\{(X_{j,1}, X_{j,2})\}_{j=1}^n$, the KDE estimate $\hat{f}(x_1, x_2)$ of its PDF $f(x_1, x_2)$ is defined as:
\begin{equation}
\hat{f}(x_1, x_2) = \frac{1}{nh_1 h_2} \sum_{j=1}^{n} K\left(\frac{x_1-X_{j,1}}{h_1}, \frac{x_2-X_{j,2}}{h_2}\right)
\end{equation}
where $n$ is the sample size, $K(\cdot, \cdot)$ is the bivariate kernel function (typically a symmetric function integrating to 1, such as a 2D Gaussian kernel), and $h_1$ and $h_2$ are the bandwidths, which control the smoothness of the estimate along each dimension. A key advantage of KDE is its independence from binning choices, and it is mathematically proven to converge to the true density faster than histograms \citep{Wasserman2006}.

In practical astronomical applications, two extensions of this standard KDE are often required, as discussed in Paper II.

(i) \textit{Weighted KDE.} When handling survey data, each data point may have a different importance, often characterized by a weight $w_j$. For example, in flux-limited surveys with complex selection functions $\mathcal{P}(z_i, L_i)$, each object is weighted by $w_j = 1/\mathcal{P}(z_j, L_j)$. The weighted KDE estimator $\hat{f}_w$ is given by:
\begin{equation}
\hat{f}_\text{w}(x_1, x_2) = \frac{1}{N_{\text{eff}} h_1 h_2} \sum_{j=1}^{n} w_j K\left(\frac{x_1-X_{j,1}}{h_1}, \frac{x_2-X_{j,2}}{h_2}\right)
\end{equation}
where $N_{\text{eff}}= \sum_{j=1}^{n} w_j$ is the effective sample size.

(ii) \textit{Adaptive KDE.} Astronomical data is often inhomogeneous, with dense clusters and sparse voids. A fixed bandwidth ( $h_1, h_2$ ) can oversmooth dense regions or undersmooth sparse regions. Adaptive KDE addresses this by allowing the bandwidth to vary for each data point $j$. The bandwidth $\lambda_j$ is made smaller in dense regions and larger in sparse regions. The adaptive KDE estimator $\hat{f}_a$ is:
\begin{equation}
\hat{f}_\text{a}(x_1, x_2) = \frac{1}{n} \sum_{j=1}^{n} \frac{1}{\lambda_{j,1} \lambda_{j,2}} K\left(\frac{x_1-X_{j,1}}{\lambda_{j,1}}, \frac{x_2-X_{j,2}}{\lambda_{j,2}}\right)
\end{equation}
where the local bandwidths $\lambda_{j,k}$ ($k=1,2$) are typically scaled from a pilot density estimate $\tilde{f}$ and global bandwidths $h_k$, such as $\lambda_{j,k} \propto h_k \tilde{f}(X_j)^{-\beta}$, with $\beta$ as the sensitivity parameter.

(iii) \textit{weighted adaptive KDE.}
In many real-world cases, such as the SDSS quasar sample analysis in Paper II, both weighting and adaptation are needed. Before applying any boundary correction, these two extensions can be combined into a weighted adaptive KDE estimator as
\begin{equation}
\hat{f}_{\rm wa}(x_1,x_2)
=
\frac{1}{N_{\rm eff}}
\sum_{j=1}^{n}
\frac{w_j}{\lambda_{j,1}\lambda_{j,2}}
K\left(
\frac{x_1-X_{j,1}}{\lambda_{j,1}},
\frac{x_2-X_{j,2}}{\lambda_{j,2}}
\right),
\label{eq:weighted_adaptive_kde}
\end{equation}
where $N_{\rm eff}=\sum_{j=1}^{n}w_j$ is the normalization factor, $w_j$ is the weight assigned to the $j$th object, and $\lambda_{j,1}$ and $\lambda_{j,2}$ are the local adaptive bandwidths along the two coordinates.

\subsection{Review: KDE-based LF Estimation for a Single Flux Limit}
\label{sec:review_paper2}

In our series of papers, the calculation of the LF $\phi(z,L)$ is abstracted as a two-dimensional density estimation problem within a specific bounded domain. Throughout this work, $L$ is defined as the logarithmic luminosity for convenience. In Paper II, we addressed the case of a single flux-limited sample, where the data domain (or survey region) $W$ is defined as:
\begin{equation}
W = \{ (z, L) \mid Z_1 < z < Z_2, L > f_{\text{lim}}(z) \}
\end{equation}
where $Z_1$ and $Z_2$ are the redshift boundaries, and $f_{\text{lim}}(z)$ is the redshift-dependent lower luminosity limit imposed by the survey's flux limit. Applying KDE directly to the domain $W$ results in significant boundary bias at the $L = f_{\text{lim}}(z)$ boundary. To solve this, we generalized the ``transformation-reflection" method in Paper II. This method involves two key steps:

\begin{enumerate}[label=(\roman*)]
    \item Transformation: We apply a variable transformation to map the bounded redshift interval $(Z_1, Z_2)$ to the unbounded space $(-\infty, \infty)$, and to transform the luminosity $L$ into a ``residual" $y$ relative to the boundary. The specific transformation is:
    \begin{equation}
    x = \ln\left(\frac{z-Z_1}{Z_2-z}\right) \quad \text{and} \quad y = L - f_{lim}(z)
    \end{equation}
    After this transformation, the data domain becomes $\{ (x, y) \mid -\infty < x < \infty, y > 0 \}$.

    \item Reflection: Although the $x$-dimension is unbounded, the $y$-dimension still has a boundary at $y=0$. To eliminate the bias from this boundary, we add ``reflection points" $(x_i, -y_i)$ about $y=0$ to compensate for the missing ``probability mass".
\end{enumerate}

Through these two steps, the original bounded data $\{ (z_i, L_i) \}$ are converted into an unbounded 2D dataset $\{ (x_i, y_i), (x_i, -y_i) \}$. We can then apply standard 2D KDE (as in \S\ref{sec:kde_basics}) to this unbounded dataset to obtain a smooth density estimate $\hat{f}(x, y)$ in the $(x, y)$ space:
\begin{eqnarray}
\label{kde_r}
\begin{aligned}
\hat{f}(x,y)=\frac{2}{2nh_1h_2}
\sum_{j=1}^{n} \left( K(\frac{x\!-\!x_j}{h_1},\frac{y\!-\!y_j}{h_2}) \!+\! K(\frac{x\!-\!x_j}{h_1},\frac{y\!+\!y_j}{h_2}) \right ).
\end{aligned}
\end{eqnarray}
Finally, $\hat{f}(x, y)$ is transformed back into the probability density $\hat{p}(z, L)$ in the original $(z, L)$ space. The luminosity function estimate $\hat{\phi}(z, L)$ is obtained:
\begin{eqnarray}
\label{phi}
\hat{\phi}(z,L)=\frac{n(Z_2-Z_1)\hat{f}(x,y|h_{1},h_{2})}{(z-Z_1)(Z_2-z)\Omega\frac{dV}{dz}},
\end{eqnarray}
where $\Omega$ is the survey solid angle and $dV/dz$ is the comoving volume element per unit solid angle.

Equation (\ref{phi}) is written for the unweighted fixed-bandwidth transformation--reflection KDE estimator. The corresponding adaptive, weighted, and weighted adaptive LF estimators can be obtained in the same way by replacing the reflected density estimate $\hat{f}(x,y)$ in Equation (\ref{kde_r}) with the corresponding KDE estimators described in Section \ref{sec:kde_basics}. Specifically, using local bandwidths yields the adaptive estimator $\hat{\phi}_{\rm a}$, using object weights yields the weighted estimator $\hat{\phi}_{\rm w}$, and combining both gives the weighted adaptive estimator $\hat{\phi}_{\rm wa}$. For the weighted versions, the normalization factor $n$ in Equation (\ref{phi}) is replaced by $N_{\rm eff}=\sum_{j=1}^{n}w_j$. More details on these estimators and their implementation are given in Paper II.

The next crucial step is to determine the optimal bandwidths ($h_1, h_2$). In Paper II, this derivation was based on the likelihood in the original $(z, L)$ space (see their Eq. 11 and 14). While correct, that approach was somewhat cumbersome. In fact, a more direct approach is to apply the Likelihood Cross-Validation (LCV) criterion directly in the unbounded, transformed $(x, y)$ space,
\begin{eqnarray}
\label{lcv1}
S_{\text{LCV}}=-2\sum_{i}^{n}&\ln[\hat{f}_{(-i)}(x_{i},y_{i}|h_1,h_2)],
\end{eqnarray}
where $\hat{f}_{(-i)}(x_i,y_i)$ is the \emph{leave-more-out} estimator as defined in Equation (13) of Paper\ II.
The optimal bandwidths $h_1,h_2$ can be obtained by numerically minimizing the objective function $S_{\text{LCV}}$.

This method proved to be efficient and accurate for samples with a single flux limit. However, when survey data consists of multiple sub-samples with different $f_{\text{lim}}(z)$, this framework requires further generalization.

\subsection{Extension to Multiple Flux-Limited Samples}
\label{sec:multiple_samples}

In real-world applications, to obtain a LF with both broad redshift coverage and a wide dynamic range in luminosity, it is common to combine datasets from different surveys (e.g., a deep, small-area survey and a shallow, wide-area survey).

\begin{figure}[!t]
	\centering
	\includegraphics[width=0.7\columnwidth]{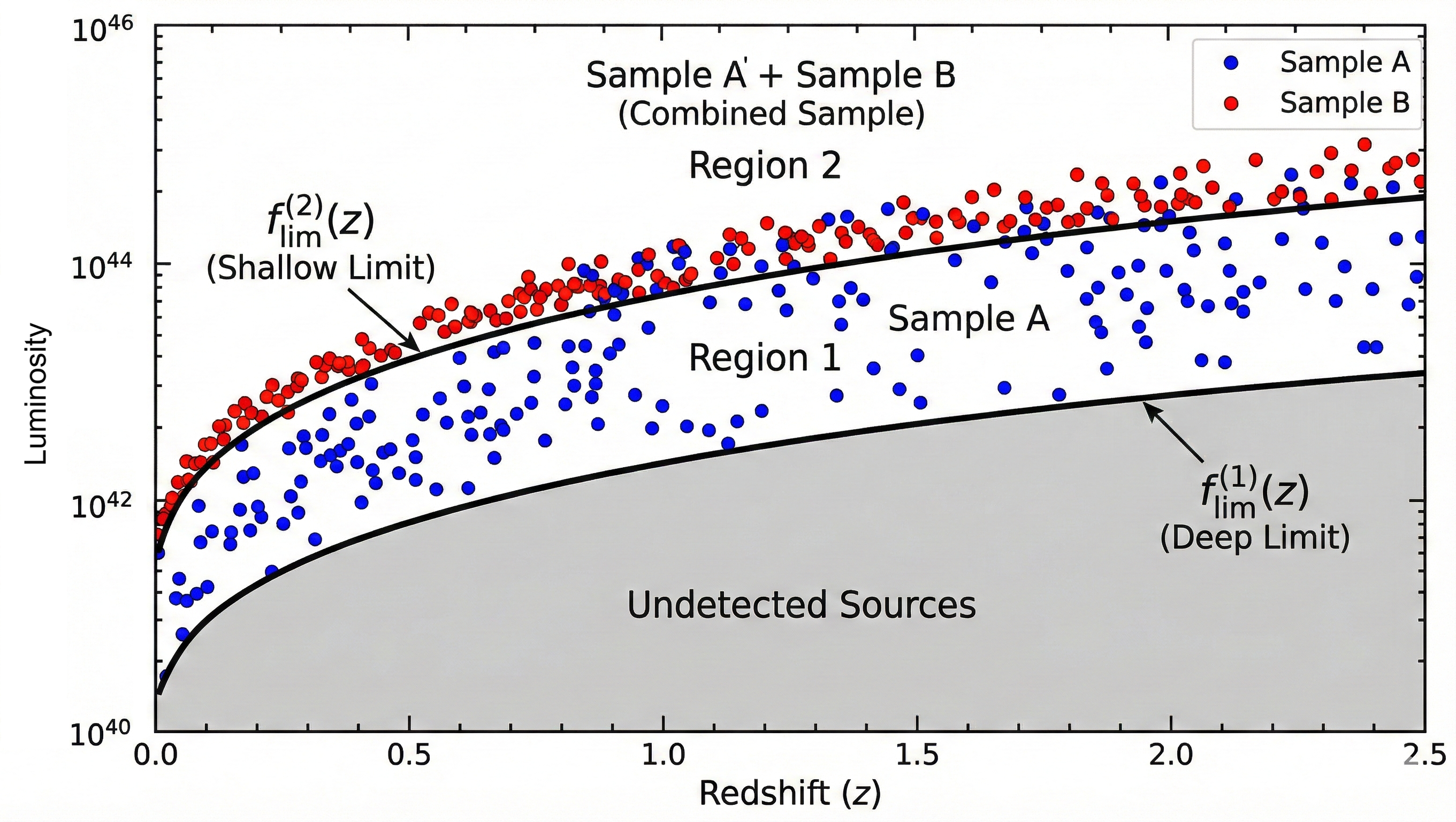}
\caption{Schematic illustration of the domain partition for the two-sample case. Sample A represents the deeper survey with a lower flux limit $f_{\mathrm{lim}}^{(1)}(z)$, while Sample B is the shallower survey with a limit $f_{\mathrm{lim}}^{(2)}(z)$. The flux curves divide the accessible $L$--$z$ plane into two disjoint regions: the faint interval ($f_{\mathrm{lim}}^{(1)} < L \le f_{\mathrm{lim}}^{(2)}$), which is populated exclusively by sources from Sample A, and the bright interval ($L > f_{\mathrm{lim}}^{(2)}$), where sources are detectable by both surveys. This segmentation forms the basis for our piecewise LF estimation.}
\label{fig:two_sample_schematic}
\end{figure}

\subsubsection{The Two-Sample Case}
Consider the case of two flux-limited samples, Sample A and Sample B, with solid angles $\Omega_1$ and $\Omega_2$, respectively. Let their flux limits correspond to redshift-dependent luminosity limits $f_{\mathrm{lim}}^{(1)}(z)$ and $f_{\mathrm{lim}}^{(2)}(z)$. Without loss of generality, we assume Sample A is the ``deeper" survey, meaning its luminosity limit is lower than that of Sample B at any given redshift:
\begin{equation}
f_{\mathrm{lim}}^{(1)}(z) < f_{\mathrm{lim}}^{(2)}(z)
\end{equation}
This configuration naturally divides the accessible $L-z$ plane into two regions. We propose a piecewise estimation strategy to integrate these heterogeneous datasets efficiently.

\textit{Step 1: Estimate for the Faint Region ($\phi_1$).}
First, we consider Sample A independently. Since Sample A covers the entire available domain down to the faint limit $f_{\mathrm{lim}}^{(1)}(z)$, we can apply the single-sample KDE method described in section \ref{sec:review_paper2} directly to Sample A. This yields an estimator $\hat{\phi}_1(z, L)$, which is valid for the entire region $L > f_{\mathrm{lim}}^{(1)}(z)$.

\textit{Step 2: Estimate for the Bright Region ($\phi_2$).}
In the region where $L > f_{\mathrm{lim}}^{(2)}(z)$, both surveys are capable of detecting sources. To reduce statistical uncertainty, we should utilize information from both samples. We construct a new, combined sample by:
\begin{enumerate}[label=(\roman*)]
    \item Taking all sources from Sample B.
    \item Extracting a subset of sources from Sample A, denoted as $A'$, which consists only of objects satisfying $L > f_{\mathrm{lim}}^{(2)}(z)$.
\end{enumerate}
Under the assumption that the contributing surveys sample the same parent population and that their selection effects are properly accounted for, the new combined sample, $S_{\text{comb}} = A' \cup B$, effectively behaves as a single homogeneous sample with a truncation boundary of $f_{\mathrm{lim}}^{(2)}(z)$. The effective solid angle for this combined sample is the sum of the individual solid angles:
\begin{equation}
\Omega_{\text{comb}} = \Omega_1 + \Omega_2
\end{equation}
We then apply the single-sample KDE method to this combined sample $S_{\text{comb}}$, using $f_{\mathrm{lim}}^{(2)}(z)$ as the reflection boundary. This yields a second estimator, $\hat{\phi}_2(z, L)$, which provides a more robust estimate for the bright end due to the larger sample volume.

\textit{Step 3: The Combined Piecewise Function.}
Finally, we integrate the results from Step 1 and Step 2 into a single LF estimate. We use $\hat{\phi}_1$ for the faint region (where only Sample A contributes) and $\hat{\phi}_2$ for the bright region (where both samples contribute):
\begin{equation}
\hat{\phi}(z, L) = \begin{cases}
\hat{\phi}_1(z, L) & \text{if } f_{\mathrm{lim}}^{(1)}(z) < L \le f_{\mathrm{lim}}^{(2)}(z) \\
\hat{\phi}_2(z, L) & \text{if } L > f_{\mathrm{lim}}^{(2)}(z)
\end{cases}
\end{equation}
It should be emphasized that this piecewise construction does not explicitly impose continuity at the transition boundaries between adjacent regions. In the ideal case where all sub-samples are drawn from the same parent population and their selection functions, flux limits, and effective survey areas are perfectly known, the adjacent estimators are expected to agree statistically at the boundary. In real applications, however, the estimators on the two sides of a boundary are optimized using different combinations of surveys and may have different effective selection functions, sample densities, and optimal bandwidths. Therefore, a small mismatch across a transition boundary can occur and should be interpreted as a diagnostic of finite-sample fluctuations and possible inter-survey systematics, rather than as a physical discontinuity in the underlying LF.

\begin{figure*}[!ht]
\centering
\includegraphics[width=0.6\columnwidth]{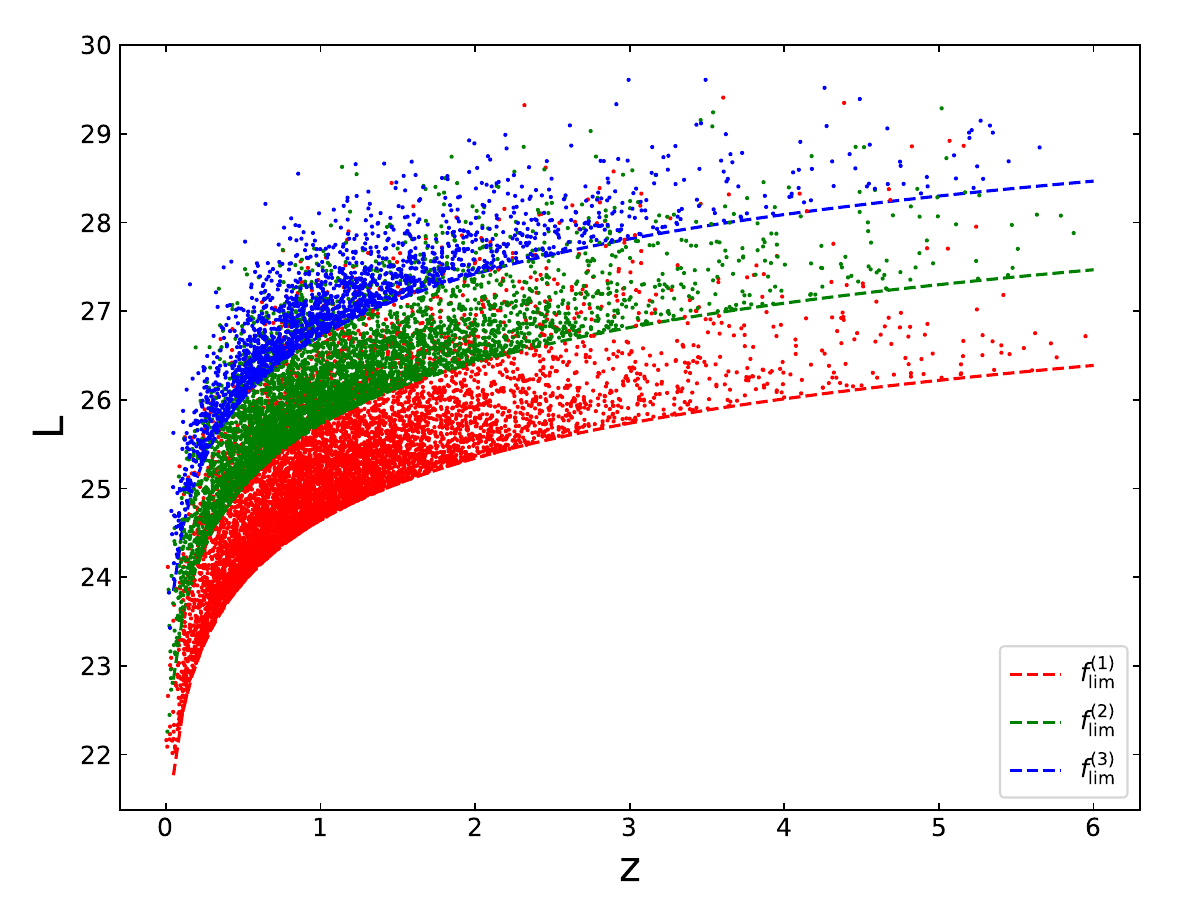}
\caption{Distribution of a representative mock realization in the luminosity--redshift ($L$--$z$) plane. The sample is composed of three disjoint survey regions with varying depths, color-coded in red ($f_{\mathrm{lim}}^{(1)}$), green ($f_{\mathrm{lim}}^{(2)}$), and blue ($f_{\mathrm{lim}}^{(3)}$), corresponding to the deepest, intermediate, and shallowest tiers, respectively. The dashed curves of the same colors indicate the corresponding luminosity limits converted from the flux detection limits. Note that the region accessible to the deepest survey (red) covers the entire domain above its flux limit, whereas the shallower surveys (green and blue) are restricted to progressively brighter luminosities.}
\label{zL_distribution}
\end{figure*}

\subsubsection{Generalization to $m$ Samples}
We now extend this strategy to the general case of $m$ flux-limited samples, denoted as $S_1, S_2, \dots, S_m$, with solid angles $\Omega_1, \Omega_2, \dots, \Omega_m$. We index these samples according to their survey depths, such that $S_1$ corresponds to the deepest survey and $S_m$ to the shallowest. Consequently, their redshift-dependent luminosity limits, $f_{\mathrm{lim}}^{(k)}(z)$, satisfy the strictly increasing order:
$$
f_{\mathrm{lim}}^{(1)}(z) < f_{\mathrm{lim}}^{(2)}(z) < \dots < f_{\mathrm{lim}}^{(m)}(z)
$$
This ordered sequence of flux-limit curves partitions the accessible $L$--$z$ plane into $m$ disjoint regions. Accordingly, the $k$-th region is defined by the interval:
$$
f_{\mathrm{lim}}^{(k)}(z) < L \le f_{\mathrm{lim}}^{(k+1)}(z)
$$
where we adopt the convention that $f_{\mathrm{lim}}^{(m+1)}(z) = \infty$. Within this interval, sources are detectable by all samples $S_j$ that are sufficiently deep, specifically those with indices $j \le k$.
For each $k \in \{1, \dots, m\}$, we construct a cumulative combined sample $S_{\text{comb}, k}$ and estimate a corresponding LF $\hat{\phi}_k(z, L)$:

\begin{enumerate}[label=(\roman*)]
    \item \textit{Sample Combination:} The combined sample $S_{\text{comb}, k}$ includes all objects from samples $S_1, \dots, S_k$ that satisfy the luminosity condition $L > f_{\mathrm{lim}}^{(k)}(z)$.
    \item \textit{Effective Solid Angle:} The effective solid angle is the sum of the solid angles of all contributing samples:
    $$
    \Omega_{\text{comb}, k} = \sum_{j=1}^{k} \Omega_j
    $$
    \item \textit{Estimation:} We apply the single-sample KDE method to $S_{\text{comb}, k}$, using $f_{\mathrm{lim}}^{(k)}(z)$ as the truncation boundary. This yields the estimator $\hat{\phi}_k(z, L)$.
\end{enumerate}

Finally, the global luminosity function $\hat{\phi}(z, L)$ is constructed as a piecewise function, utilizing the estimator $\hat{\phi}_k$ within its optimal validity range:
$$
\hat{\phi}(z, L) = \hat{\phi}_k(z, L) \quad \text{for } f_{\mathrm{lim}}^{(k)}(z) < L \le f_{\mathrm{lim}}^{(k+1)}(z)
$$
where we define $f_{\mathrm{lim}}^{(m+1)}(z) \to \infty$. This generalized framework ensures that for any luminosity $L$, the LF estimate utilizes the maximum available number of galaxies from all survey regions capable of detecting objects at that luminosity, weighted by their total sky coverage.

It is useful to emphasize that the multi-sample problem is not solved by introducing a fundamentally new KDE estimator. Instead, the key step is to decompose the original multi-flux-limit problem into a sequence of piecewise single-boundary problems. For each luminosity region defined by two adjacent flux-limit curves, we construct the appropriate combined sample, adopt the corresponding local truncation boundary, and use the effective solid angle of the contributing surveys. Once this reduction has been made, the LF estimation within each region can be carried out using the single-flux-limit KDE estimators developed in Paper II. These include the full bivariate transformation--reflection KDE estimator, its adaptive and weighted extensions, and the one-dimensional small-sample approximation together with its adaptive version. Therefore, the choice of estimator in each piecewise region should be made according to the properties of the corresponding combined sample. In particular, when the local combined sample is small, the small-sample approximation of Paper II  (Section 2.5) can be applied within that region, and the relevant sample size is the number of objects in that piecewise combined sample rather than the total number of objects in all surveys.
In practice, each of these piecewise single-boundary calculations can be carried out directly with the public \texttt{kdeLF} Python package developed in our previous work (Paper II).\footnote{\url{https://github.com/yuanzunli/kdeLF}; archived version: \doi{10.5281/zenodo.6033776}.}, by supplying the corresponding combined sample, local truncation boundary, and effective solid angle.

\begin{figure*}[!htb]
\centering
\includegraphics[width=1.00\columnwidth]{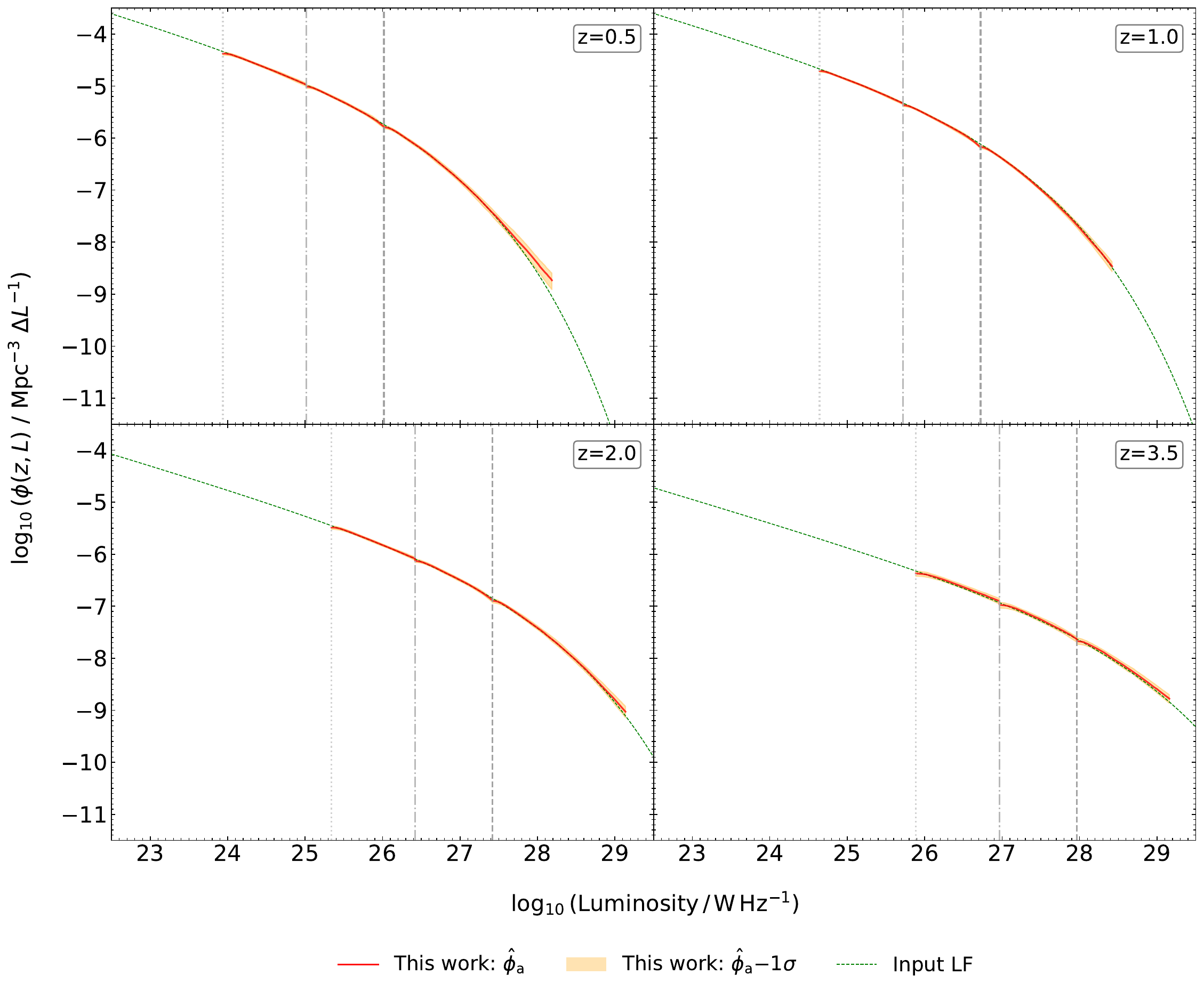}
\caption{Median LFs and uncertainties derived from 200 simulated realizations using our generalized KDE method (specifically the adaptive estimator, $\hat{\phi}_{\mathrm{a}}$). The panels show the results at four representative redshifts: $z=0.5, 1.0, 2.0$, and $3.5$. The red solid curves represent the median of the 200 estimates, while the orange shaded regions indicate the $1\,\sigma$ dispersion across the 200 samples. The green dashed curves show the ground-truth input LF (Model A from \citet{2017ApJ...846...78Y}). The vertical gray lines (dotted, dashed-dotted, and dashed) mark the positions of the flux limits $f_{\mathrm{lim}}^{(1)}$, $f_{\mathrm{lim}}^{(2)}$, and $f_{\mathrm{lim}}^{(3)}$, respectively, corresponding to the three nested survey tiers.}
\label{fig:median_LF}
\end{figure*}

\section{results}
\subsection{Application of KDE to Simulated Samples}
\label{sec:Simulated_Samples}

To rigorously evaluate the performance of our generalized KDE method, we applied it to a series of mock catalogs generated from a known underlying population. We adopted the parameterized Radio Luminosity Function (RLF) from \citet{2017ApJ...846...78Y} (Model A) as the input model. The simulations cover a redshift range of $0.0 \le z \le 6.0$. All sources are assumed to follow a power-law spectrum ($S \propto \nu^{-\alpha}$) with a spectral index of $\alpha = 0.75$.

We simulated a survey strategy comprising three disjoint sky regions with different depths and areas. The adopted flux limits ($F_{\mathrm{lim}}$) and corresponding solid angles ($\Omega$) are: (1) $1\,\mathrm{mJy}$ over $0.0132\,\mathrm{sr}$, (2) $12\,\mathrm{mJy}$ over $0.0453\,\mathrm{sr}$, and (3) $120\,\mathrm{mJy}$ over $0.1404\,\mathrm{sr}$. These configurations yielded sample sizes of $n_{\mathrm{data}}^{(1)} = 8010$, $n_{\mathrm{data}}^{(2)} = 5010$, and $n_{\mathrm{data}}^{(3)} = 2010$, respectively, resulting in a total of 15,030 sources. Based on this setup, 200 independent Monte Carlo realizations were generated to quantify the statistical uncertainties.

Figure \ref{zL_distribution} illustrates the $L$--$z$ distribution for a representative mock realization. The three sub-surveys are color-coded in red, green, and blue, representing the deep, intermediate, and shallow fields, respectively. Their corresponding flux limits are marked by dashed lines. This visualization highlights the tiered observational coverage: the faintest regime (between the red and green boundaries) is sampled solely by the deepest survey, while the brightest regime (above the blue boundary) is covered by the combined solid angle of all three surveys.

We applied the generalized multi-sample framework proposed in Section \ref{sec:multiple_samples} to the ensemble of 200 mock catalogs. Given the nested structure of the three sub-surveys with different flux limits, we adopted a piecewise estimation approach to construct the global LF.
For the density estimation within each valid region, we employed the adaptive estimator, $\hat{\phi}_a$, provided in our \texttt{kdeLF} Python package. In the present multi-sample application, \texttt{kdeLF} is applied separately to the corresponding piecewise combined sample, using the appropriate local flux limit and effective solid angle.

The statistical performance of our method is summarized in Figure \ref{fig:median_LF}. The red solid curves depict the median LFs derived from the 200 realizations, while the orange shaded areas represent the $1\,\sigma$ dispersion across the 200 samples. Across all representative redshift snapshots ($z=0.5, 1.0, 2.0$, and $3.5$), the median estimates exhibit excellent agreement with the input reference LF across the entire dynamic range. Furthermore, the remarkably narrow dispersion indicates that the adaptive estimator possesses high precision and robustness, effectively suppressing the statistical noise inherent in individual realizations even at the boundaries where different survey depths overlap.

\begin{figure*}[!ht]
\centering
\includegraphics[width=0.8\columnwidth]{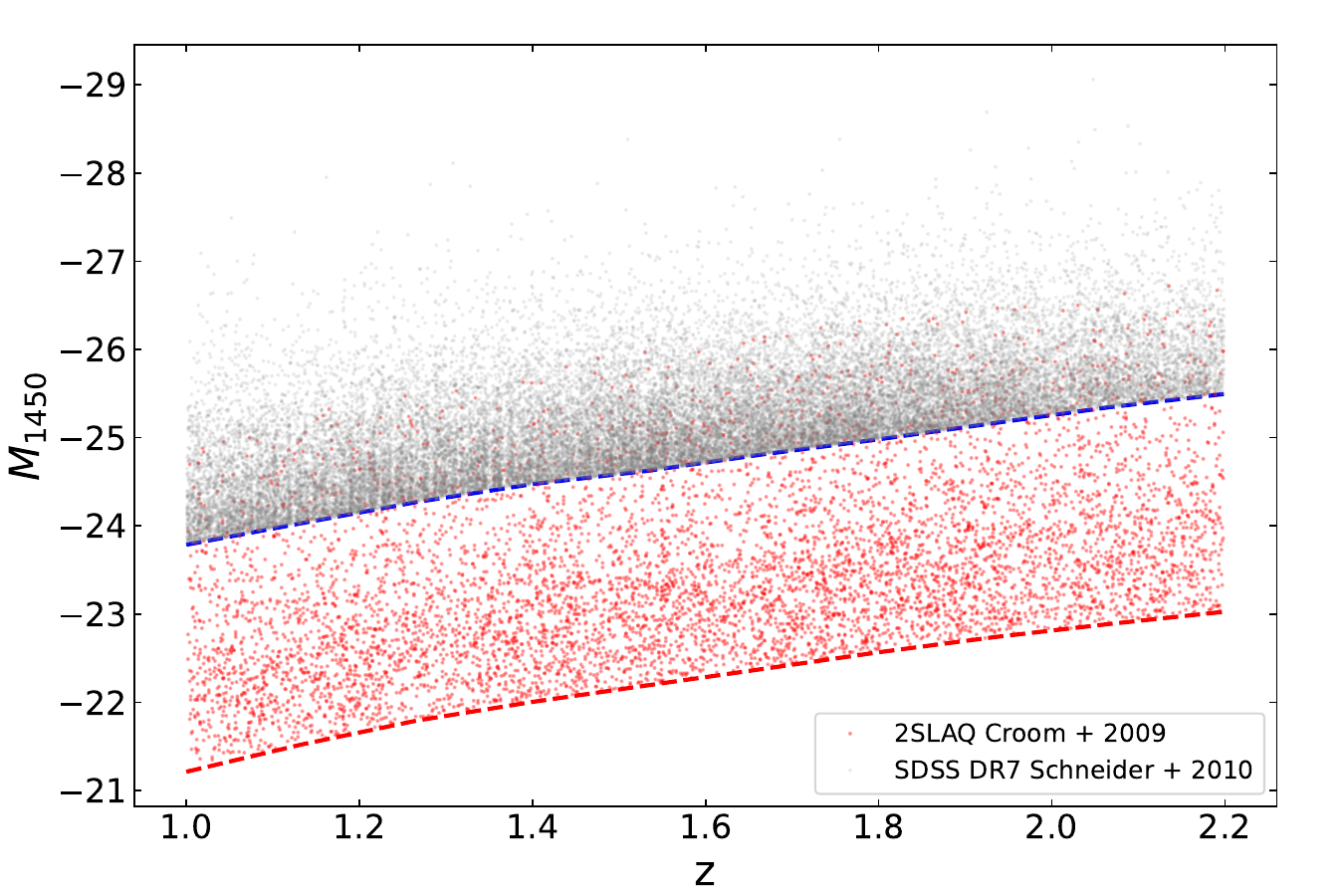}
\caption{Distribution of the quasar sample compiled by \citet{2019MNRAS.488.1035K} in the $M_{1450}$--$z$ plane, restricted to $1.0 \le z < 2.2$. Red points show the SDSS DR7 subsample (32{,}548 quasars; wide area, relatively shallow), while green points indicate the 2SLAQ subsample (7{,}090 quasars; narrower area, deeper). The dashed lines indicate the respective flux-limit curves for the two surveys. This visualization highlights the tiered coverage of the datasets, where the two samples overlap in $M_{1450}$ while sampling different comoving volumes.}
\label{fig:Mz_distribution}
\end{figure*}

\begin{figure*}[!ht]
\centering
\includegraphics[width=1.02\columnwidth]{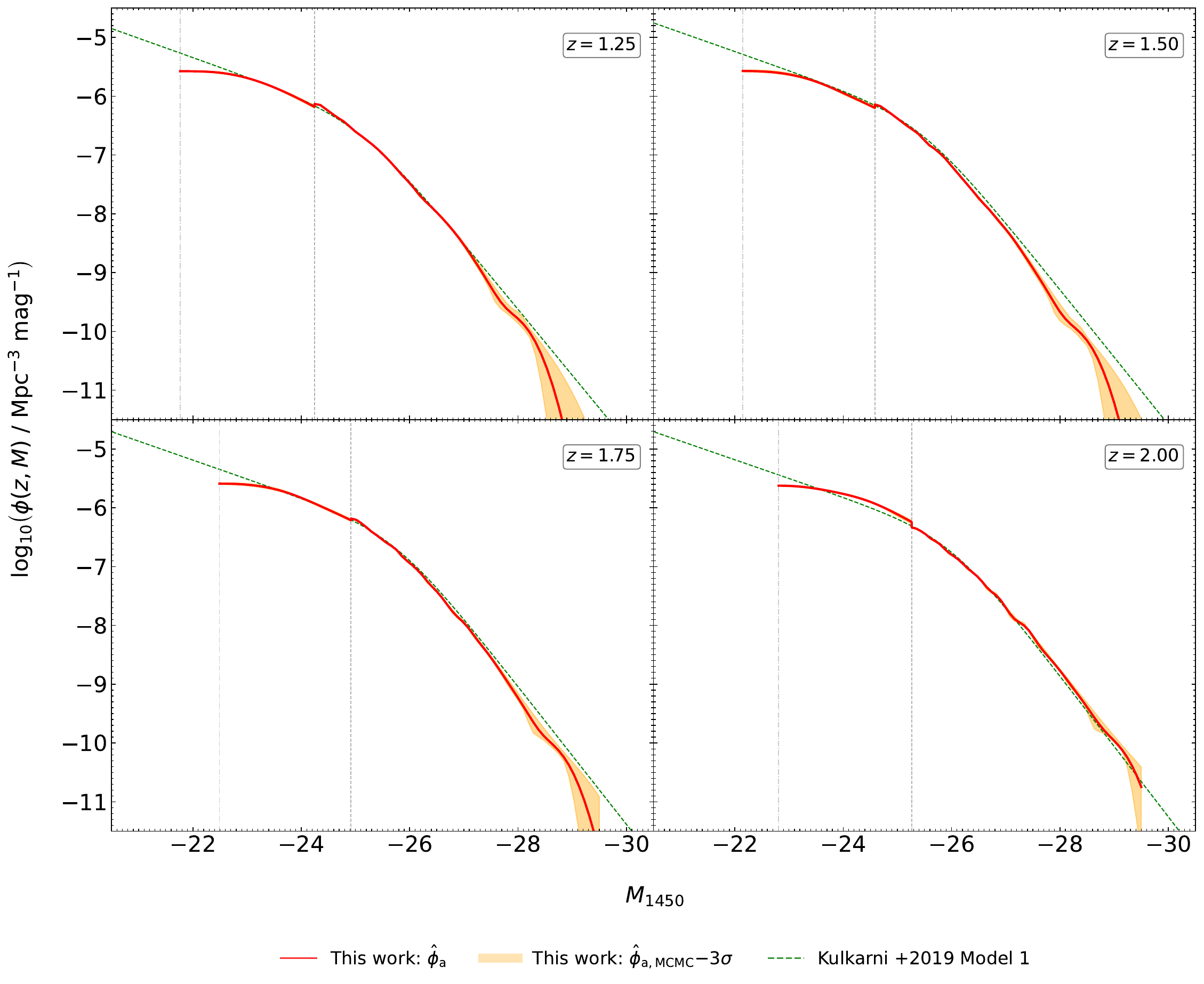}
\caption{Quasar UV LF estimates at four representative redshifts within $1.0 < z < 2.2$. The red solid curves represent our fiducial generalized KDE estimates (using the $\hat{\phi}_{\mathrm{a}}$ estimator), with orange shaded regions indicating the $3\sigma$ credible intervals derived from MCMC analysis. The green dashed lines denote the reference LF from \citet[][Model 1]{2019MNRAS.488.1035K}. Vertical grey dash-dotted and dashed lines mark the luminosity limits of the 2SLAQ and SDSS DR7 surveys, respectively, highlighting the piecewise boundaries of the tiered observational data.}
\label{mcmc_akdeLF}
\end{figure*}


\subsection{Application of KDE to Real quasar Survey Data}
\label{quasar_data}

In this section, we apply our generalized KDE method to real observational data from the homogenized quasar sample compiled by \citet{2019MNRAS.488.1035K}. This catalog unifies multiple optical surveys by standardizing the cosmology and converting fluxes to absolute monochromatic AB magnitudes at a rest-frame wavelength of 1450\,\AA\ ($M_{1450}$). Specifically, we analyze the subsamples derived from the SDSS DR7 quasar catalog \citep{2010AJ....139.2360S} and the 2SLAQ survey \citep{2009MNRAS.392...19C}, which provide complementary coverage of the bright and faint ends of the luminosity function, respectively.

We restrict our analysis to the redshift range $1.0 \le z < 2.2$. While \citet{2019MNRAS.488.1035K} excluded sources at $z < 0.6$ to mitigate uncertainties arising from host galaxy light correction and extended source incompleteness, we adopt a more conservative lower limit of $z=1.0$. Since \citet{2019MNRAS.488.1035K} noted that host galaxy contamination becomes negligible at $z > 0.8$, our cutoff at $z=1.0$ ensures that the sample is strictly dominated by point sources and free from residual host galaxy systematics. The upper limit is set to $z=2.2$, consistent with the boundaries of the low-redshift analysis in \citet{2019MNRAS.488.1035K}.

Figure \ref{fig:Mz_distribution} illustrates the distribution of the selected quasars in the $M_{1450}$--$z$ plane. The red points denote the SDSS DR7 sample, comprising 32,548 quasars and providing high-luminosity coverage over a large survey area of $6248~\mathrm{deg}^{2}$. Conversely, the green points represent the 2SLAQ sample, which includes 7,090 quasars reaching significantly fainter magnitudes in deeper but narrower fields that comprise the North Galactic Pole (NGP, $127.7~\mathrm{deg}^{2}$) and the South Galactic Pole (SGP, $64.2~\mathrm{deg}^{2}$). The corresponding flux limits for each survey are indicated by the dashed lines. This tiered data structure demonstrates the necessity of our multi-sample framework, as the two surveys overlap in luminosity while probing different comoving volumes.

We then apply our generalized KDE framework to this tiered sample by implementing the piecewise estimation strategy described in Section \ref{sec:multiple_samples}. In each valid region, we perform the density estimation using the adaptive estimator, $\hat{\phi}_{\mathrm{a}}$, provided in the \texttt{kdeLF} Python package (Paper II).
The resulting UV LFs at four representative redshifts are presented in Figure \ref{mcmc_akdeLF}. These estimates are shown as red solid curves, representing the median of the posterior distribution, with the orange shaded regions indicating the $3\sigma$ credible intervals derived from our Markov Chain Monte Carlo (MCMC) analysis. For comparison, we plot the reference UV LF (Model 1) from \citet{2019MNRAS.488.1035K} as green dashed lines.

Our generalized KDE method shows excellent agreement with the reference model across the majority of the luminosity range. This consistency validates our approach, particularly in how the adaptive bandwidth manages the varying sample density across the tiered datasets. At the bright end ($M_{1450} \lesssim -28$), where the data are relatively sparse, the adaptive kernels provide a stable reconstruction of the LF tail, mitigating the unphysical fluctuations often seen in non-adaptive methods.

A careful inspection of Figure \ref{mcmc_akdeLF} shows that, although the overall agreement with the reference model is good, a small discontinuity is visible near the transition boundary where the contributing survey set changes. This feature should not be interpreted as a physical jump in the quasar LF. It mainly reflects the piecewise nature of the estimator: the two sides of the boundary are inferred from different effective samples and are optimized independently. Therefore, our present implementation does not mathematically enforce equality of the adjacent estimates at the boundary.

The fact that the offset can be larger than the formal MCMC dispersion is also understandable. The MCMC uncertainty shown here mainly reflects the uncertainty of the KDE smoothing parameters under the adopted input sample and flux-limit functions. It does not fully include possible survey-to-survey systematics, such as residual differences in selection criteria, completeness corrections, photometric calibration, effective survey area, or field-to-field variance between SDSS DR7 and 2SLAQ. The boundary offset therefore provides a useful diagnostic of such inter-survey systematics and finite-sample effects. A more complete treatment would require propagating survey-dependent selection functions, completeness uncertainties, and possible cross-normalization uncertainties into the LF reconstruction. We do not impose an explicit continuity constraint in the present demonstration, because such a constraint could artificially hide real inconsistencies between heterogeneous surveys.

\begin{figure*}[!htb]
\includegraphics[height=8cm,width=8cm]{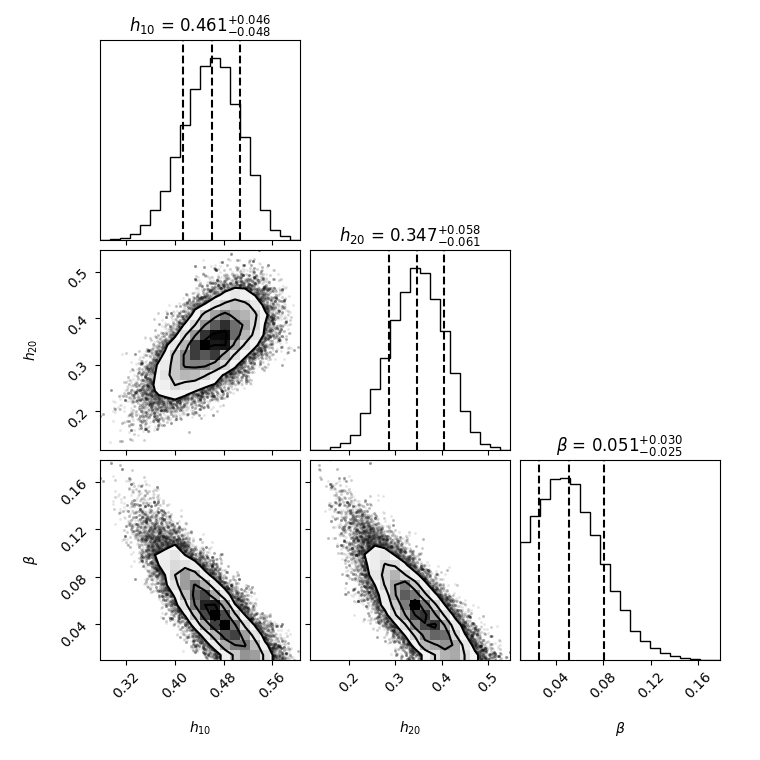}  \qquad
\includegraphics[height=8cm,width=8cm]{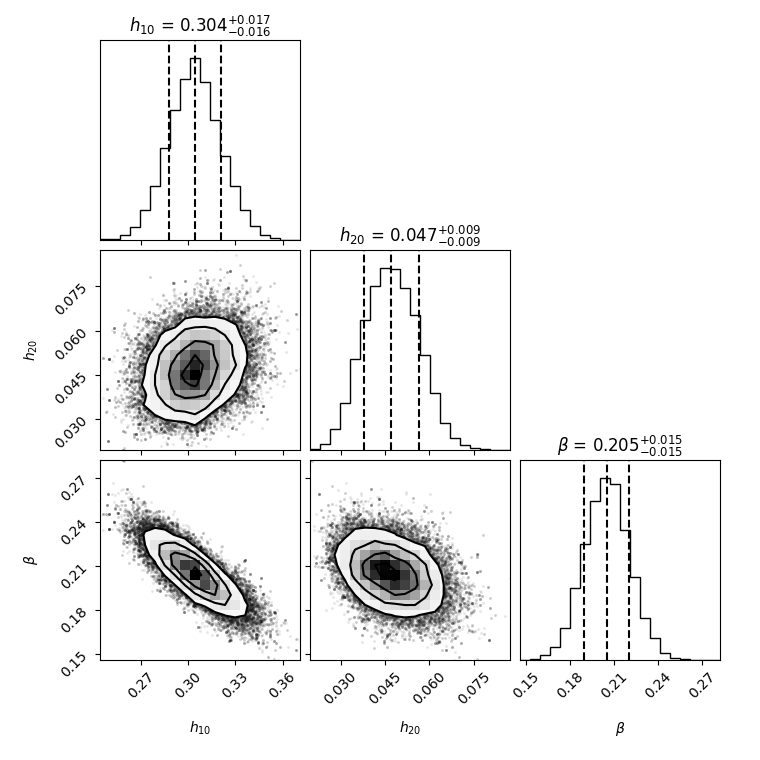}
\caption{Posterior distributions of the adaptive KDE parameters derived using the \texttt{kdeLF} package. Left and right panels correspond to the first and second steps of our piecewise estimation, respectively.}
\label{fig_mcmc}
\end{figure*}

The optimal parameters for our adaptive KDE estimator are determined using the MCMC module integrated within the \texttt{kdeLF} package, which utilizes the \texttt{emcee} sampler \citep{2013PASP..125..306F}. Following our piecewise strategy, we conduct two independent MCMC runs. The resulting posterior distributions for the three adaptive KDE parameters—the global bandwidths and the sensitivity parameter—are presented in Figure \ref{fig_mcmc}. The left panel displays the corner plot for the first step, focusing on the faint region ($f_{\mathrm{lim}}^{(1)} < L \le f_{\mathrm{lim}}^{(2)}$) where only 2SLAQ data are available. The right panel shows the posterior distributions for the second step, where parameters are independently optimized for the bright region ($L > f_{\mathrm{lim}}^{(2)}$) using the combined SDSS and 2SLAQ sample. In both cases, the well-converged chains and clear posterior peaks demonstrate that the parameters are statistically robust and provide a reliable basis for the final LF reconstruction.


\section{Discussion and Conclusions}
\label{sec:conclusion}

In this work, we have presented a significant generalization of our kernel density estimation (KDE) framework for luminosity function (LF) estimation, extending its applicability from single flux-limited samples to multiple, heterogeneous datasets. By partitioning the luminosity-redshift ($L-z$) plane based on varying flux limits and adopting a piecewise estimation strategy, our method effectively integrates information from disjoint sky regions with different observational depths.

The robustness of this generalized approach was first validated through 200 independent Monte Carlo simulations. The results demonstrate that the median LF estimates are in excellent agreement with the ground-truth input across a wide range of luminosities and redshifts. Notably, the narrow dispersion across the realizations indicates that the adaptive KDE estimator is highly precise and effectively suppresses statistical noise, even in sparse regions or near survey boundaries where different depths overlap.

Furthermore, the application of our method to the real quasar survey data compiled by \citet{2019MNRAS.488.1035K}, combining the SDSS DR7 and 2SLAQ subsamples, further confirms its practical utility. The derived UV LFs are broadly consistent with existing parametric models and provide a flexible, model-independent reconstruction of the quasar population. Compared with traditional binned estimators, the KDE framework avoids arbitrary bin choices and suppresses many binning-induced fluctuations. However, a small residual discontinuity is visible near the transition boundary where the contributing survey set changes. We interpret this offset not as a physical feature of the LF, but as a diagnostic of the independent piecewise estimation and possible inter-survey systematics, such as residual differences in selection criteria, completeness corrections, effective survey area, or field-to-field variance.

This piecewise reduction also clarifies how the method should be used for much smaller samples, such as high-redshift JWST galaxy samples distributed over disjoint fields \citep[e.g.,][]{Adams2024,Whitler2025}. In such cases, the total number of sources in all surveys is not the quantity that controls the performance of the estimator. Instead, one must consider the number of objects available in each piecewise combined sample. For a region with sufficient statistics, the full bivariate transformation-reflection KDE or its adaptive version is appropriate. For a region containing only a small number of sources, the one-dimensional small-sample approximation developed in Paper II provides a more stable coarse-grained estimate. The resulting LF should therefore be interpreted as a smoothed non-parametric reconstruction rather than a high-resolution measurement of local LF structure. This approach avoids arbitrary luminosity binning and accounts for different survey depths and sky areas, but it cannot remove the intrinsic Poisson noise, completeness uncertainties, photometric-redshift uncertainties, or field-to-field cosmic variance associated with very small samples. Bootstrap or Monte Carlo resampling and comparison with binned or parametric estimates remain useful complementary checks.

The key advantages and contributions of our generalized KDE method are summarized as follows:
\begin{enumerate}[label=(\roman*)]
    \item Flexibility and Accuracy: By abstracting the LF calculation as a 2D density estimation problem within a bounded domain, we successfully utilize the ``transformation-reflection'' technique to mitigate boundary biases, ensuring the reconstruction is physically consistent near the flux limit.

    \item Optimal Resource Utilization: The piecewise framework ensures that for any given luminosity, the LF estimate incorporates the maximum available number of sources from all contributing surveys, effectively maximizing the statistical power of multi-tier survey designs.

    \item Adaptive Smoothing: The integration of adaptive bandwidths provides a stable and smooth reconstruction within each piecewise region, particularly suppressing unphysical fluctuations at the bright end where data points are typically sparse.

    \item Algorithmic Maturity: The KDE estimation in each piecewise region is performed using our previously developed public Python package \texttt{kdeLF}, which was introduced in Paper II. No separate KDE estimator is required for the multi-sample case: the present framework is implemented by constructing the appropriate combined sample for each flux-limit region and then applying the existing \texttt{kdeLF} estimator with the corresponding local truncation boundary and effective solid angle.

\end{enumerate}

As modern extragalactic astronomy enters the era of massive, multi-tiered surveys---such as the Square Kilometre Array (SKA), the Legacy Survey of Space and Time (LSST), and the Dark Energy Survey (DES)---the ability to self-consistently combine datasets of varying sensitivities becomes paramount. To meet the computational demands of these upcoming surveys, future iterations of our framework could incorporate fast KDE algorithms, such as those based on Fast Fourier Transforms \citep[e.g.,][]{Gramacki2017,DB2018}, to significantly reduce the computational cost while maintaining non-parametric flexibility. Future developments will focus on further extending this method to include more complex selection functions and multi-wavelength completeness corrections. Such advancements will further establish our KDE-based framework as a highly competitive alternative to existing nonparametric estimators, characterized by its superior accuracy and statistical robustness in reconstructing the evolution of cosmic populations.

\begin{acknowledgments}
We thank the anonymous reviewer for the many constructive comments and suggestions, leading to a clearer description of these results. We acknowledge the financial support from Science Fund for Distinguished Young Scholars of Hunan Province (grant No. 2024JJ2040), and the National Natural Science Foundation of China (grant No. 12073069). Z.Y. is supported by the Xiaoxiang Scholars Programme of Hunan Normal University.
\end{acknowledgments}

%

\vspace{5mm}
\software{SciPy \citep{2020NatMe..17..261V}, Astropy \citep{2013A&A...558A..33A},
emcee \citep{2013PASP..125..306F}, corner \citep{ForemanMackey2016}, GetDist \citep{2019arXiv191013970L}, QUADPACK \citep{1983qspa.book.....P}, matplotlib \citep{2007CSE.....9...90H}, \texttt{kdeLF} \citep{2022ApJS..260...10Y}.}









\end{document}